# Host-guest co-amorphous structure revealed by the suppression of the first sharp diffraction peak in isotactic poly(4-methyl-1-pentene)

Tomoki Ogihara,[1] Yusuke Hiejima,[2] and Ayano Chiba[1*]

[1]Department of Physics, Keio University, Yokohama, 223-8522, Japan
[2]Department of Chemical and Materials Science, Kanazawa University, 920-1192, Japan

**ABSTRACT.** While host-guest co-crystals are well established, and co-amorphous solids are recognized in materials science, the concept of a "host-guest co-amorphous" structure remains largely unexplored. A potential analogue is seen in $SiO_2$ glass under high pressure with helium as a pressure medium; the significant drop in compressibility in this system is ascribed to helium atoms occupying internal voids. In this study, we investigated a semicrystalline polymer, isotactic poly(4-methyl-pentene-1) (P4MP1), which shares key characteristics with $SiO_2$ glass, particularly regarding the first sharp diffraction peak (FSDP). The FSDP in P4MP1 is attributed to internal voids, as evidenced by its suppression under pressure and recovery upon decompression for molten P4MP1. Notably, the response to helium as a pressure medium is also known to parallel the behavior observed in $SiO_2$ glass. Here, we analyzed two-dimensional X-ray diffraction (2D-XRD) patterns of stretched P4MP1 and found a suppression of FSDP when P4MP1 is immersed in decane. The use of stretched samples enabled the clear isolation of the amorphous FSDP from overlapping crystalline diffractions. Our findings reveal the existence of a host-guest co-amorphous system at room temperature and atmospheric pressure, in which decane molecules occupy the amorphous host matrix of P4MP1. Unlike conventional co-amorphous mixtures, this structure is defined by the specific accommodation of guests within the host's inherent voids. Intriguingly, the signature of this structure in diffraction measurements, manifested as changes in the FSDP intensity ratio, may be regarded to parallel the variations in Bragg peak intensity ratios in host-guest co-crystals. Since selective sorption and guest exchange are well-known in co-crystals, hosts capable of forming co-amorphous structures will be promising materials for molecular sieves, or more generally, liquid-phase molecular sieves.

## I. INTRODUCTION

Host-guest co-crystals, in which guest molecules are arranged within a crystalline host matrix, are well established. Similarly, co-amorphous solids, characterized by the random mixing of two or more components, are also recognized in the field of materials science. However, a structure that can be defined as a "host-guest co-amorphous" system has yet to be formally proposed, in which guest molecules are incorporated into a solid amorphous host matrix. To date, the most closely related phenomenon has been reported in the field of $SiO_2$ glass under high pressure; specifically, a drastic reduction in compressibility occurs when helium is used as a pressure medium, a behavior attributed to helium atoms occupying the internal voids [1,2]. However, there are no prior reports of this structure being obtained at ambient conditions. Although equivalent structures might have appeared in polymer research, they have not been formally identified as this specific type of host-guest system.

If a host-guest co-amorphous structure exists, how would the X-ray diffraction patterns change before and after guest occlusion? In the case of well-studied host-guest co-crystals, there are examples where the symmetry of the host crystal remains unaltered even after removal of the guest molecules [3]. In such instances, the intensity ratio of Bragg peaks change before and after the guest occlusion. Even if new diffraction peaks appear to emerge upon guest occlusion, they represent a redistribution of intensity ratios, where previously negligible reflections become observable due to the modified structure factor. By analogy, if guest molecules are incorporated into an amorphous host without significantly altering the host's disordered framework, one might expect a corresponding change in the peak intensity ratios. The present study clarifies that this is indeed the case: the intensity ratio of the first sharp diffraction peak changes upon guest occlusion.

The first peak of the structure factor $S(Q)$ in a wide range of structurally disordered systems is referred to as the first sharp diffraction peak (FSDP). This peak, typically observed in the low-$Q$ region (where $Q$ denotes the momentum transfer), is broadly regarded as a signature of intermediate-range order [4-6].

As a prototypical example, the FSDP of $SiO_2$ glass is recognized and has been suggested to be associated with the presence of voids in the network structure [7, 8]. Indeed, $SiO_2$ glass is known to exhibit substantial structural transformations upon compression, characterized by a dramatic reduction in the FSDP intensity [9, 10]. Recently, angstrom-beam electron diffraction has enabled direct and detailed observations showing that the FSDP of $SiO_2$ glass originates from atomic density fluctuations, specifically voids [11]. There is additional evidence that the FSDP of $SiO_2$ glass originates from voids: when helium is used as a pressure medium, the reduction in FSDP intensity is suppressed because the helium atoms penetrate these voids during compression [1, 2].

It is well-established that the FSDP serves as the most sensitive indicator of structural rearrangement during pressure-induced polyamorphic structural changes, as it originates from the largest-scale structural voids that are inherently highly pressure-sensitive. Specifically, long-range structural features, such as dihedral angles defined by four atomic positions, exhibit greater flexibility than short-range features such as bond lengths consisting of two atoms, provided that sufficient void space is available.

In 2012 we proposed that polymer systems can also exhibit a peak that could be termed as FSDP [12]. The first peak in the structure factor $S(Q)$ of polymers is frequently referred to as the "interchain peak." Historically, it has been termed the "polymerization peak" because it emerges as a new feature at the lowest scattering angles upon the polymerization of monomers [13]. This peak is also variously known as the "intermolecular peak," "larger than van der Waals peak" [14], "inter-main-chain peak" [15], or "amorphous halo". In the case of isotactic poly(4-methyl-1-pentene) (P4MP1) [16, 17] shown in Fig. 1, this peak is observed at approximately 0.7 Å$^{-1}$ [18]. This peak shares key characteristics with the typical of FSDP of amorphous materials such as $SiO_2$ glass, in two primary respects. First, the interchain peak of P4MP1 is most likely to reflect the presence of structural voids [19]. Second, P4MP1 exhibits behavior suggestive of pressure-induced polyamorphism [12]. Below, we elaborate on these similarities in further detail, and hereafter we shall call the interchain peak of P4MP1 the FSDP.

The presence of bulky branched side chains makes P4MP1 the lightest among all commercially available polymers. Specifically, P4MP1 is noted for its unusually low crystalline density of 0.83 g/cm$^3$, while the density of its amorphous state is nearly identical to that of the crystalline state at room temperature [16, 17, 20]. This exceptional lightness, alongside its robust chemical resistance, represents one of the most distinctive features of P4MP1. Such low density implies the existence of significant structural void spaces; indeed, the permeability reported for gases such as $CO_2$ and $CH_4$ [21] further supports the presence of these internal voids within the material.

In Ref. 12, we reported that the X-ray diffraction profile of molten P4MP1 exhibits a pronounced decrease in the intensity of the FSDP with increasing pressure, followed by its recovery upon decompression. Notably, the FSDP remains relatively unchanged when helium gas is employed as a pressure medium [19], presumably because helium atoms penetrate into the interstitial voids. These behaviors are strikingly analogous to those observed in $SiO_2$ glass.

Such a reduction in the FSDP intensity under pressure can be attributed to a structural change from a loosely-packed to a densely-packed structure, namely polyamorphism, in molten P4MP1. The presence of a maximum in the pressure dependence of the melting point, along with the reported occurrence of pressure-induced amorphization [22], supports the existence of polyamorphism in P4MP1.

These parallels with $SiO_2$ glass support a void-based origin for the FSDP. In this paper, we show that alkanes, specifically decane, can penetrate the voids in the amorphous region of solid P4MP1 at room temperature, resulting in a reduction in FSDP intensity. In other words, we show that the emergence of a host-guest co-amorphous structure is manifested as a reduction in FSDP intensity.

Semicrystalline polymers, such as P4MP1, consist of coexisting crystalline and amorphous regions, as the macromolecular nature of the chains inherently precludes complete crystallization. To investigate the behavior of the FSDP—a characteristic feature of disordered systems—it is essential to distinguish the scattering contributions from each region. In this study, we employ uniaxially stretched P4MP1 sheets, as such oriented samples allow us to effectively decouple the X-ray diffraction signals of the crystalline and amorphous components.

*Contact author: ayano@phys.keio.ac.jp

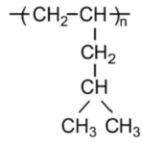

FIG 1. Molecular structure of poly(4-methyl-1-penten) (P4MP1).

## II. METHODS

### A. Materials and Preparation

Uniaxially stretched samples with a thickness of 80 μm were prepared following the procedure described in Ref. 23. The crystallinity of the stretched sample was 38 % by X-ray diffraction measurements.

Decane was employed as the solvent for molecular occlusion. Reagent-grade normal decane was purchased from Kanto Chemical Co. Inc. For neutron diffraction, fully deuterated decane, with the degree of deuterium substitution 99 atom % , was purchased from ISOTEC. These solvents were used without further purification.

### B. X-ray diffraction

X-ray diffraction (XRD) measurements were conducted at BL40B2, SPring-8 (Japan) [24], using an incident energy of 22 keV and a camera length of 476.98 mm. Pilatus3S 2M was used as the detector. All measurements were performed at room temperature.

After a measurement of a dry stretched sample, decane was dropped on and the measurement of the wet sample was done 7 hours later since several hours may be required for the solvent to fully penetrate the stretched samples and reach equilibrium [23]. Measurements were also performed 20 minutes after solvent immersion, and the results are provided in the Supplemental Material [25]. The samples were folded with 5-μm-thick Kapton films for the measurements. Further details can be found in our previous paper [23]. The background scattering, including the contribution from the Kapton films, was appropriately subtracted.

### C. Small-angle neutron scattering

Small-Angle Neutron Scattering (SANS) experiment was performed with TAIKAN (BL15) at J-PARC (Japan) [26]. A stretched P4MP1 sample was mounted in a cell with $SiO_2$ glass windows. Following the initial dry state measurement, deuterated decane was introduced into the cell to measure the solvent-sorbed state. All data were collected at room temperature,

*Contact author: ayano@phys.keio.ac.jp

with background subtractions (including the cell contribution) performed accordingly.

## III. RESULTS AND DISCUSSIONS

### A. X-ray diffraction

Figure 2 shows a two-dimensional XRD pattern of the dry stretched sample. The image consists of a superposition of spot-like Bragg reflections and a ring-shaped amorphous halo, due to coexistence of crystalline and amorphous regions. In more detail, the XRD image exhibits (i) a ring-shaped amorphous FSDP, (ii) the spot-like Bragg reflections such as the 200 reflection, (iii) a horizontal streak, and (iv) a secondary amorphous halo.

The amorphous FSDP and the crystalline 200 reflection appear at the same wavenumber, making them indistinguishable in isotropic samples. However, by using a stretched sample, we successfully isolated the FSDP to investigate its structural evolution during solvent sorption. Namely, the FSDP observed at 0.67 $Å^{-1}$ almost overlaps with the 200 Bragg reflection (0.67 $Å^{-1}$), calculated from the P4MP1 Form I crystal structure (space group $P\underline{4}b2$ ; $a = 18.70$ Å [27]).

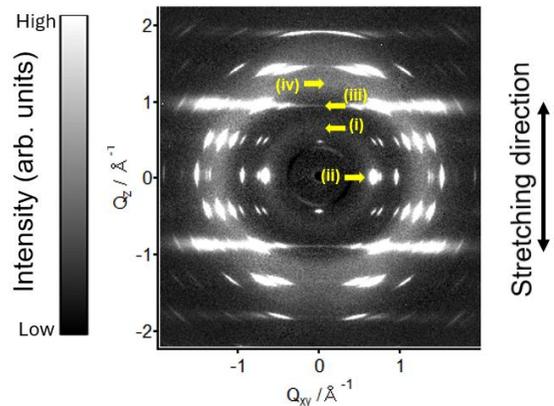

FIG 2. Two-dimensional XRD pattern of stretched P4MP1: (i) amorphous FSDP (ring), (ii) crystalline 200 Bragg reflection (spot), (iii) horizontal streak, and (iv) secondary amorphous halo. The arrow on the far right indicates the stretching direction of the sample. The 200 reflection is located at a wavenumber almost overlapping with the FSDP.

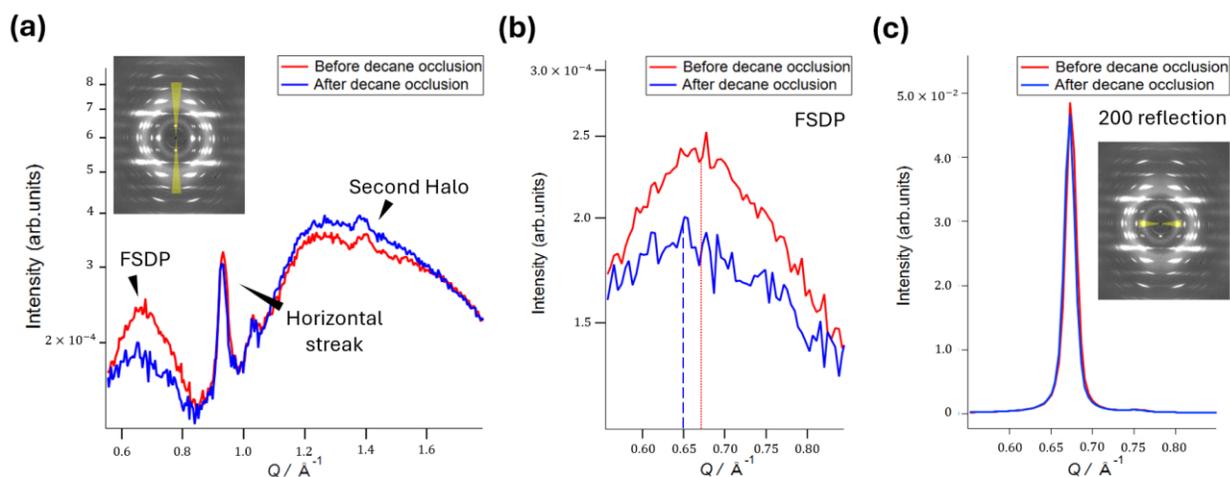

FIG 3. X-ray diffraction profiles of stretched P4MP1 before (denoted by red lines) and after (blue lines) decane occlusion. (a) Diffraction profiles in wide $Q$ range. The yellow-shaded parts in the inset show the azimuthal averaging range used to obtain the profiles. Only meridional sectors are used to avoid crystalline reflections and thus the profile mainly reflects the structure of the amorphous regions. (b) Magnified view of the FSDP region. The dotted and dashed lines indicate the peak positions before and after solvent occlusion, respectively. (c) Diffraction profiles of the 200 reflections. Only equatorial sectors (the small yellow-shaded parts in the inset) are used. More comprehensive data, including the original two-dimensional diffraction images and their time-dependent changes, are available in the Supplemental Material [25].

The changes in the FSDP upon solvent sorption are presented in Fig. 3. Plots in (a) and (b) represent the diffraction profiles integrated over the meridional sector, as highlighted in yellow in the inset of Fig. 3(a), which allows for the isolation of the amorphous component from the equatorial crystalline reflections. The highlighted sector corresponds to azimuthal angles of 82.5°–92.5° and 262.5°–277.5°, where the horizontal direction in inset is defined as 0°. Figure 3(a) shows the FSDP at 0.67 Å$^{-1}$ and the second halo at 1.3 Å$^{-1}$. The peak observed around 0.92 Å$^{-1}$ corresponds to a portion of the horizontal streak from the crystalline region as seen in Fig. 2(iii).

The FSDP intensity decreases upon decane occlusion [Fig. 3(a)(b)]. Considering that the FSDP originates from the scattering contrast between the polymer backbones and the intervening voids, these results provide direct evidence for the formation of a host-guest co-amorphous structure. The occlusion of decane into these voids reduces this contrast, leading to the observed drop in FSDP intensity. Since both P4MP1 and decane consist solely of carbon and hydrogen, carbon acts as the primary X-ray scatterer; thus, the FSDP intensity directly reflects the presence of voids—essentially regions where carbon atoms are absent. Therefore, the observed intensity decrease upon solvent occlusion demonstrates the formation of a host-guest co-amorphous system, provided that decane is absorbed into the amorphous regions of the stretched P4MP1. The preferential occlusion into the amorphous region is further supported by small-angle neutron scattering (SANS) measurements, as detailed in the following subsection.

Upon decane occlusion, a slight shift in the FSDP position from 0.67 Å$^{-1}$ to 0.65 Å$^{-1}$ was also observed. These values were determined by Gaussian fitting. The respective peak positions are indicated by the dotted and dashed lines in Fig. 3(b). While the FSDP intensity reflects the amount of voids, the peak position will reflect the interchain distance. Therefore, the peak shift indicates that the average distance between the main chains in the amorphous region expands by approximately 3% due to solvent sorption in the voids. The assignment of the FSDP peak position to the interchain distance can be inferred from the interchain distance in the crystalline phase; the lattice constant $a = 18.70$ Å and the space group $P\bar{4}b2$ implies an interchain distance of approximately 9 Å, which matches the value derived from the FSDP position ($2\pi/0.7$Å$^{-1} \sim 9$ Å). Furthermore, since the length of the repeating unit is constrained to less than twice the C-C bond length (1.5 Å), the structural origin of the sharp FSDP in the amorphous region is unlikely to be the side chains. Instead, it is more appropriately

*Contact author: ayano@phys.keio.ac.jp

attributed to the characteristic distance between the main chains [12].

It is worth noting that the 200 reflection shifted from 0.674 Å$^{-1}$ to 0.672 Å$^{-1}$ upon decane occlusion [Fig. 3(c), calculated using the azimuthal angle ranges of -8.5°–8.5° and 171.5°–188.5°]. This shift corresponds to only 0.3%, indicating that the shift observed in the FSDP of the amorphous regions is more pronounced. Since both the 200 reflection and the FSDP reflect the backbone-backbone distance, the lower flexibility of the crystalline regions compared to the amorphous regions is consistent with the fact that the amorphous regions primarily occlude the solvent, as will be experimentally verified by SANS in the next subsection.

### B. Small-angle neutron scattering

We also conducted SANS experiments to ensure the decane occlusion into the amorphous region of the stretched P4MP1 sample. Fully deuterated decane was used to reduce the background level caused by the high incoherent scattering cross-section of hydrogen.

Figure 4(a) shows two-dimensional SANS patterns of oriented P4MP1 samples before (left) and after (right) the occlusion of deuterated decane. As seen in the left image, the scattering intensity of the dry sample is notably low, indicating a lack of scattering contrast. The low scattering intensity is also evident from the red line in Fig. 4(b), which represents the azimuthal average of the 2D pattern. This low intensity is due to the fact that the densities of the crystalline and amorphous regions in P4MP1 are nearly identical; consequently, the lamellar periodic peak, typically observed in semicrystalline polymers due to the contrast between these two phases, is not detected.

Upon deuterated decane occlusion, a pair of meridional arcs appears along the stretching direction, which provides the evidence for the solvent occlusion in the amorphous region. We note that, as is typical for semicrystalline polymers, molecular occlusion occurs predominantly within the amorphous regions rather than the crystalline domains [23]. As shown in Fig. 4(b), these arcs correspond to a broad peak at $Q = 0.016$ Å$^{-1}$ in the one-dimensional SANS profile [blue line in Fig. 4(b)]. This peak is attributed to the long period (long spacing) of the lamellar crystals [28]. While the emergence of a lamellar peak upon solvent occlusion has been previously demonstrated for isotropic P4MP1 [23], the present study provides the first measurement of this phenomenon in oriented samples. In contrast to the isotropic ring observed in unoriented samples, the scattering in the stretched sample appears as meridional sectors. This result indicates that the lamellar stacks are predominantly oriented parallel to the stretching axis. We may note that the difference in the baseline level in the high-$Q$ region between the red and blue lines is attributed to the presence of excess bulk deuterated decane.

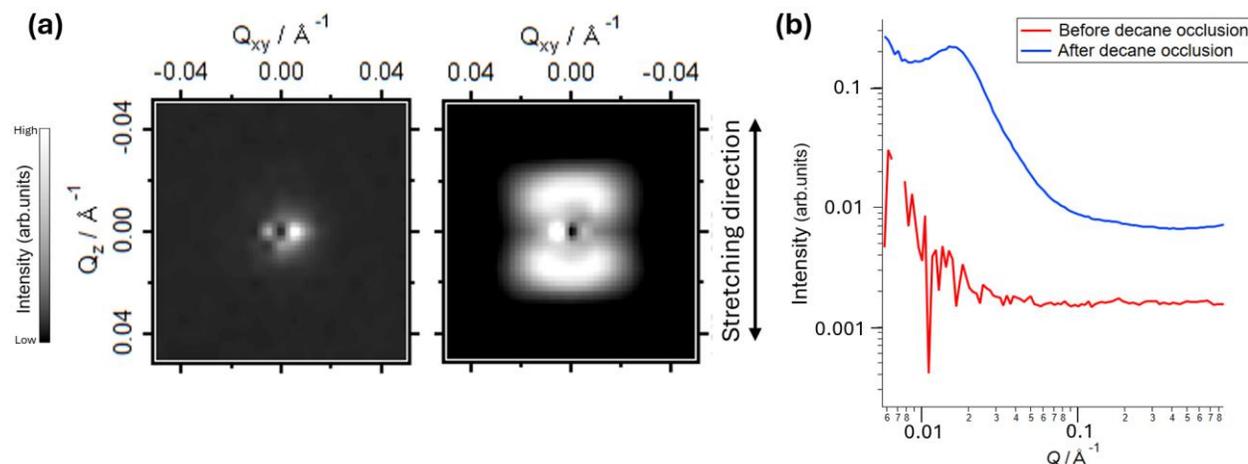

FIG 4. SANS patterns of oriented P4MP1 before and after deuterated decane occlusion. (a) Two-dimensional patterns before and after occlusion are shown as left and right figures, respectively. (b) The SANS profiles obtained by azimuthal averaging. The profiles before and after occlusion are shown as red and blue lines, respectively. The profile of dry sample (red line) appears noisy due to the lack of scattering contrast, as the densities of the crystalline and amorphous regions are nearly identical for P4MP1.

*Contact author: ayano@phys.keio.ac.jp

These two results stand in contrast despite both being obtained from stretched samples: the FSDP appears as an isotropic ring at approximately 0.7 Å$^{-1}$ [Fig. 2 (XRD)], while the lamellar peak manifests as a pair of meridional sectors at approximately 0.016 Å$^{-1}$ [Fig. 4 (SANS)]. These indicate that while the lamellar structure, composed of crystalline and amorphous regions in oriented P4MP1, is clearly organized along the stretching axis, the voids within the amorphous regions possess no specific orientation. This discrepancy is reasonable given that Å-scale voids could be isotropic in a macroscopically stretched sample. Nevertheless, the overall isotropy of the amorphous regions, i.e., whether it remains truly isotropic at Å-scale, within the stretched semicrystalline polymer remains a subject for future investigation.

From the perspective of the universality of the FSDP, it is particularly noteworthy that the FSDP in P4MP1 originates from voids significantly larger than those in SiO$_2$ and GeO$_2$. While the microscopic voids in these oxide glasses are sized to allow the penetration of He (∼ 2.6 Å in diameter) but exclude H$_2$ (∼ 2.9 Å) [2], the voids in P4MP1 are sufficiently large to accommodate much larger molecules, such as decane.

It is also worth noting the possibility that the effects of stretching can be partially similar to those of compression. The FSDP of P4MP1 shown in Fig. 3 may not appear as sharp as the well-known FSDP profiles in other systems. However, this is an effect of the uniaxial orientation of the sample; in contrast, the FSDP of molten P4MP1 is intrinsically sharp [12]. During uniaxial stretching, the direction perpendicular to the tensile axis experiences compressive stress. Therefore, the amorphous regions of oriented P4MP1 may effectively correspond to a state in which amorphous P4MP1 has been pressurized and subsequently returned to ambient pressure. Investigating the relationship between the pressure history of amorphous P4MP1 and the behavior of its FSDP remains a subject for future study.

Finally, two additional similarities between P4MP1, SiO$_2$, and H$_2$O, beyond those related to FSDP and polyamorphism, are presented below. First, tetrahedral local motifs have been linked to the polyamorphism in systems such as H$_2$O and SiO$_2$. While P4MP1 may not appear to possess such structures at first glance, we may point out that every carbon atom in the polymer chain forms a tetrahedral geometry via sp$^3$ hybridized orbitals. Consequently, the structural behavior of P4MP1 can be effectively reduced to a packing problem of interconnected tetrahedra.

Second, the presence of void spaces, characterized by FSDP, may facilitate the formation of host-guest structures. Just as water molecules sustain voids through hydrogen bonding to form methane hydrates, or as helium atoms occupy the internal voids of SiO$_2$ glass, decane molecules can enter the voids created by the bulky side chains of P4MP1, leading to the formation of a host-guest co-amorphous structure.

The formation of host-guest structure naturally leads to the usage as molecular sieves. Just like the hosts in co-crystals, amorphous hosts will function as a molecular sieve through guest selectivity and guest exchange. Indeed, P4MP1 exhibits selective occlusion of longer alkanes from mixtures of long- and short-chain alkanes [29]. Based on the present results and those in Ref. 23, it is probable that this selectivity originates within the amorphous region, i.e., the guest selectivity of host-guest co-amorphous.

The reason P4MP1 can act as a host in co-amorphous systems is that its bulky side chains maintain the voids even after the removal of guest molecules. Most notably, P4MP1 exhibits a pronounced FSDP in the liquid state, suggesting the formation of a porous liquid structure [19] inherent to its molecular architecture, which is very simple as shown in Fig. 1. Leveraging this unique feature could open up new possibilities for the design of liquid molecular sieves.

### IV. CONCLUSIONS

We investigated FSDP of P4MP1 before and after immersion in decane. As a result, a decrease in the intensity of FSDP was observed, accompanying the occlusion of decane into the amorphous region. The loss of scattering contrast upon decane molecules filling the FSDP-related voids demonstrates the formation of a host-guest co-amorphous system.

Specifically, by using uniaxially oriented P4MP1 samples, we were able to distinguish the diffraction of the amorphous region from that of the crystalline region. The halo-like FSDP observed on the two-dimensional detector exhibited a decrease in intensity upon the occlusion of decane, accompanied by a shift in the peak position toward lower angles by approximately 3%. The intensity of the FSDP in P4MP1 reflects the total volume of voids, while its peak position corresponds to the backbone-backbone distance. Therefore, the observed decrease in the intensity indicates the occlusion in the amorphous region whereas the shift in the peak position indicates that the incorporation of solvent molecules led to an expansion of the backbone-backbone distance.

Furthermore, small-angle neutron scattering (SANS) measurements were performed to confirm that the solvent is predominantly occluded within the amorphous region. For these measurements, deuterated decane was employed as the solvent. Upon the occlusion of the deuterated decane, a lamellar peak of P4MP1 emerged along the stretching direction. This is because the similar densities of the crystalline and amorphous phases in P4MP1 prevent the observation of a lamellar peak in the dry sample and decane occlusion to the amorphous region results in the observable scattering contrast between them. Furthermore, the fact that this lamellar peak appeared along the stretching direction indicates that the alternating structure of crystalline and amorphous phases is oriented parallel to the tensile axis. Combining the XRD and SANS results, it is clarified that in oriented P4MP1, the lamellar periods possess a clear structural orientation along the stretching direction, whereas voids within the interlaminar amorphous regions show almost no directional preference.

*Contact author: ayano@phys.keio.ac.jp

In summary, we report the formation of a host-guest co-amorphous structure, characterized by a decrease in FSDP intensity. Such a response in diffraction is analogous to the intensity variations known in host-guest co-crystals. While the current study observes a decrease in FSDP intensity due to the comparable scattering powers of the host and guest, an increase in intensity is theoretically possible if guests with significantly different scattering powers are employed.

ACKNOWLEDGMENTS

The authors thank Hiromi Murashige (KU Leuven, Belgium) for collaboration on the measurement, Noboru Ohta (SPring-8) and Hiroki Iwase (CROSS) for their valuable support at BL40B2 and TAIKAN, respectively. We also thank Rintaro Inoue and Ken Morishima (Kyoto Univ.) for their extensive support during the preliminary SANS experiments at SANS-U (JRR-3, Japan). This work was supported by KAKENHI 21H05574, 22K03558, and by the Asahi Glass Foundation. Experiment numbers are 2022B1538 (Spring-8), 2022B0343 (J-PARC), and 22561 (JRR-3). We also thank Yoshiharu Nishiyama (CNRS, France) and Ryo Akiyama (Kyushu Univ.) for the valuable discussion.

*Contact author: ayano@phys.keio.ac.jp

*Contact author: ayano@phys.keio.ac.jp

# Supplemental Material for
# "Host-guest co-amorphous structure revealed by the suppression of the first sharp diffraction peak in isotactic poly(4-methyl-1-pentene)"

Tomoki Ogihara,[1] Yusuke Hiejima,[2] and Ayano Chiba[1*]

[1]Department of Physics, Keio University, Yokohama, 223-8522, Japan
[2]Department of Chemical and Materials Science, Kanazawa University, 920-1192, Japan

In this Supplemental Material, we present the two-dimensional diffraction patterns that provided the basis for Fig. 3 in the main text. These images demonstrate that the intensity of the circular FSDP decreases in a largely isotropic manner upon solvent occlusion.

Figure S1 shows the two-dimensional diffraction patterns of (a) the dry stretched P4MP1, (b) 20 minutes after immersion in decane, and (c) 7 hours after immersion. The horizontal gap (the thick black line) near the beam center is due to the space between the two-dimensional detector modules. In Fig. 2 of the main text, this gap was eliminated by merging data collected at three different detector positions. For (b) only, the scattering from bulk decane was measured separately and subtracted; however, a residual contribution remains visible as a high-intensity ring-like feature on the right side of the image (b). Due to these analytical inconsistencies regarding the bulk decane subtraction, these specific data were not included in the main text.

While a sharp FSDP, labeled as (i), is observed in (a), it appears broader and weaker in (b) and (c). It is observed that the FSDP intensity has already decreased after 20 minutes, as shown in (b). (b) and (c) show the peak intensity weakens almost isotopically, accompanied by a broadening of the profile.

It is also noteworthy that the 001 Bragg peak, labeled as (ii) in (a), exhibits a decrease in intensity upon solvent occlusion. Although this reflection should ideally be absent due to extinction rules, its actual observation has been pointed out since the 1960s [S1]. As shown in Figs. S1(b) and (c), the intensity of the 001 reflection gradually diminishes. This attenuation suggests that crystalline disorder is reduced, tending toward elimination as the solvent is occluded; in other words, a guest-induced ordering of the crystal lattice likely occurs. A more detailed discussion of this behavior will be presented in a separate report. Regarding solvent-induced changes in the crystalline reflections, the variation was most prominent in the 001 peak. Other changes were limited to subtle shifts in diffraction positions, such as that of the 200 reflection described in the main text. No changes in the intensity ratio were observed except for the 001 reflection.

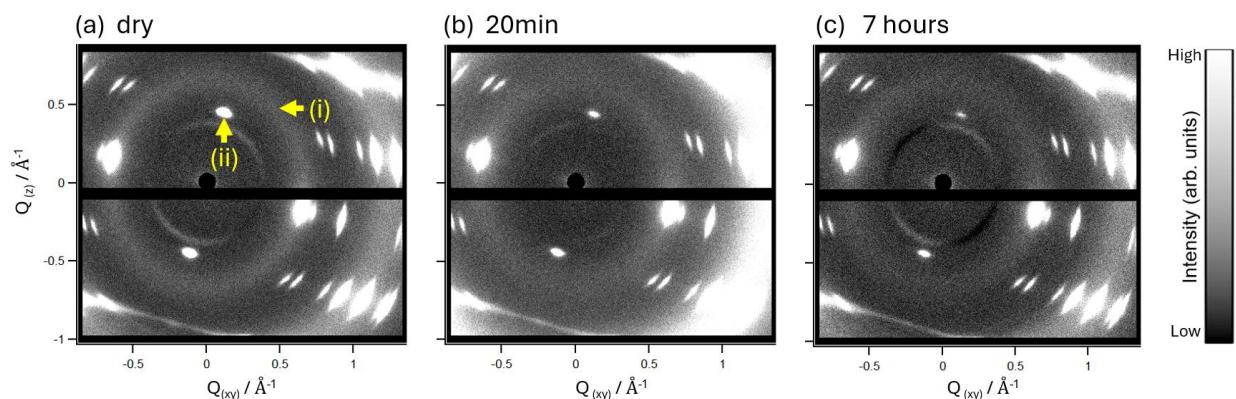

FIG S1. Two-dimensional X-ray diffraction patterns of (a) dry stretched P4MP1, and the same sample at (b) 20 minutes and (c) 7 hours after immersion in decane. In (a), the circular profile (i) is the FSDP from the amorphous region, and the spot (ii) is the 001 reflection from the crystalline regions. Only for (b), the data are presented with the bulk decane contribution subtracted, as the original signal included significant scattering from the excess solvent.

[S1] M. Litt, J. Polym. Sci., Part A, 1, 2219 (1963).